\documentclass[twocolumn,english,superscriptaddress]{revtex4}
\usepackage[latin9]{inputenc}
\usepackage{amsmath}
\usepackage{amssymb}
\usepackage{graphicx}
\usepackage{color}

\makeatletter

\@ifundefined{textcolor}{}
{%
 \definecolor{BLACK}{gray}{0}
 \definecolor{WHITE}{gray}{1}
 \definecolor{RED}{rgb}{1,0,0}
 \definecolor{GREEN}{rgb}{0,1,0}
 \definecolor{BLUE}{rgb}{0,0,1}
 \definecolor{CYAN}{cmyk}{1,0,0,0}
 \definecolor{MAGENTA}{cmyk}{0,1,0,0}
 \definecolor{YELLOW}{cmyk}{0,0,1,0}
}

\usepackage{babel}

\usepackage{babel}

\makeatother

\usepackage{babel}
\begin{document}

\title{{\normalsize{}{}{}{}{}{}{}High voltage assisted mechanical
stabilization of single-molecule junctions}}

\author{{\normalsize{}{}{}{}{}{}{}David Gelbwaser-Klimovsky}}

\affiliation{{}Department of Chemistry and Chemical Biology, Harvard University,
Cambridge, MA 02138}

\author{{\normalsize{}{}{}{}{}{}{}Al\'an Aspuru-Guzik}}

\affiliation{{}Department of Chemistry and Chemical Biology, Harvard University,
Cambridge, MA 02138}

\author{{\normalsize{}{}{}{}{}{}{}Michael Thoss}}

\affiliation{Institut f\"ur Theoretische Physik and Interdisziplin\"ares Zentrum f\"ur Molekulare Materialien, Friedrich-Alexander-Universit\"at Erlangen-N\"urnberg, Staudtstr.\,7/B2, D-91058 Erlangen, Germany}

\author{{\normalsize{}{}{}{}{}{}{}Uri Peskin}}

\affiliation{{}Schulich Faculty of Chemistry, Technion-Israel Institute of Technology,
Haifa 32000, Israel}
\begin{abstract}
{\normalsize{}{}{}{}{}{}{}The realization of molecular-based
electronic devices depends to a large extent on the ability to mechanically
stabilize the involved molecular bonds, while making use of efficient resonant
charge transport through the device. Resonant charge transport can
induce vibrational instability of molecular bonds, leading to bond
rupture under a bias voltage. In this work, we go beyond the wide-band approximation in order to study the phenomenon of vibrational
instability in single molecule junctions and show that the energy-dependence
of realistic molecule-leads couplings affects the mechanical stability
of the junction. We show that the chemical bonds can be stabilized
in the resonant transport regime by increasing the bias voltage
on the junction. This research provides guidelines for the design
of mechanically stable molecular devices operating in the regime of
resonant charge transport. }{\normalsize \par}
\end{abstract}
\maketitle
Chemical bond rupture is a major concern when single molecules are
being considered as electronic components in nano-scale devices \cite{hybertsen2017modeling,tao-dna,latha-si,latharupture}.
In single molecule junctions, tunneling electrons temporally dwell
on the molecule and therefore induce changes in the molecular charging
state. In the deep (or off-resonant) tunneling regime charge fluctuations
on the molecule during transport lead to energy exchange between the
electronic and the mechanical molecular degrees of freedom \cite{natelson-prl,galperin2004inelastic,caspary2007electronic,smit2002measurement}.
These processes have remarkable effect on the molecular junction transport
properties, but their influence on the mechanical stability of chemical
bonds is considered to be minor. {}However, resonant tunneling, often
associated with relatively high bias voltage, is more relevant for
electronics than deep tunneling, since the associated currents are
significantly larger. In this regime, changes in the charging state
of the molecule are pronounced, and consequently the electronic coupling
to molecular vibrations can result in bond rupture either at the
molecule or at the molecule-lead contacts.{} This mechanical instability
often limits experiments on single molecule junctions to the off-resonant
tunneling regime. In order to combine the desired features of efficient
resonant transport at high voltage operation with mechanically stable
molecules, one needs to determine which experimentally controlled
parameters contribute to the mechanical stability of molecules under
non-equilibrium transport conditions.

{}Charge transport induced bond rupture was observed for physisorbed
molecules in scanning tunneling microscope experiments \cite{huang2013single,huang2014vibrational,ho}
as well as in atomic chains \cite{sabater2015evidence} and single
molecule junctions \cite{latha-si,latharupture,capozzi2016mapping},
where the molecules are chemically bonded to the leads. In particular,
the occurrence of bond rupture increased with increasing bias voltage, which points to the increased transport induced charging
of the molecule. It is worthwhile to mention in the present context
that the possibility to control bond rupture by the molecular junction
parameters (e.g., voltage, coupling to the leads, etc) is relevant
not only for the sake of mechanical stability of nano-scale current
carrying devices, but also for nano-scale chemical catalysis. It was
shown theoretically that (by a proper design) transport induced heating
can be directed towards a particular bond \cite{hartle2010communication,volkovich2011bias},
suggesting the possibility of mode-selective chemistry in single junction
architectures.

{}Theoretical works on bond dissociation induced by resonant tunneling
through molecular junctions consider the effective (anharmonic) mechanical
force on the nuclei when the electronic state is a mixture of different
charging states \cite{kosov1,kosov2,kochmorse}. This may turn the
bound nuclear geometry into a metastable one, leading to bond rupture
in the steady state (long-time) limit. {}Other theoretical approaches
restrict the discussion of molecular vibration excitations to the
harmonic approximation \cite{kochmorse,hartle2011vibrational,hartle2015effect,volkovich2011bias,hartle2010communication}{}.
While bond dissociation can not be treated explicitly in this case,
the occurrence of vibrational instability \cite{kochmorse,hartle2011vibrational,hartle2015effect,kast2011charge,avriller2009electron} due to the excess of energy flow into vibrations is considered as the indicator
for bond rupture in the anharmonic case.

{}In this work we address one of the crucial aspects of the realization
of single molecule electronic devices: How can the conditions of operation
be tuned in order to benefit from efficient resonant charge transport
at high voltage through a single molecule junction, and yet to maintain
the mechanical stability of the molecule? For this purpose we consider
in detail the generic model of vibrational heating in non-equilibrium
transport between two Fermionic reservoirs. The onset of vibrational
instability is analyzed in the limit of weak molecule-lead and intra-molecular
vibronic couplings, where resonant charge transport kinetics is expressed
in terms of vibrational heating and cooling processes. We demonstrate
cases where increasing the bias voltage favors cooling processes over
heating, thus stabilizing the molecular junction at a higher voltage.
This result contrasts with the common intuition for resonant transport,
which correlates instability with higher voltage. However, it is readily
explained by considering realistic, energy-dependent, profiles for
the density of states in the leads, beyond the commonly invoked wide-band approximation. Indeed, relative changes in the leads densities
of states may favor inelastic transport of low-energy electrons from
one lead into high energy states of the other leads, resulting in
efficient vibrational cooling. Since the relevant densities of states
depend on the bias voltage, this effect can be obtained at relatively
high voltages. This analysis provides new guidelines for mechanical
stabilization of single molecule junctions under resonant transport
conditions.

{}The  minimal model for transport induced vibrational excitation
considers a single electronic transport channel through the molecule \cite{mitra2004phonon},
where a single spin orbital is coupled to a single bond, represented
as a quantum mechanical oscillator. In realistic systems vibrational
excitation energies exceeding a few ($\sim$ 10-100) vibration quanta
would be typically associated with highly anharmonic parts of the
potential energy surface, where bond rupture is likely to occur. However,
our purpose is to capture the onset of vibrational instability at
low vibration excitation numbers, consistent with the harmonic approximation.
Therefore we shall treat explicitly the harmonic part of the potential
energy surface. The model Hamiltonian reads, $H=[\hbar\tilde{\omega}_{0}+\frac{\hbar g}{2}(\tilde{a}+\tilde{a}{}^{+})]\tilde{d}^{\dagger}\tilde{d}+\hbar\Omega\tilde{a}{}^{\dagger}\tilde{a},$
where $\hbar\tilde{\omega}_{0}$ is the charging energy of the single
molecular orbital, associated with the Fermionic creation and annihilation 
operators, $\tilde{d}^{\dagger}$ and $\tilde{d}$, respectively,
and $\tilde{a}^{\dagger}$ ($\tilde{a})$ are the creation (annihilation)
operators for the vibrational degree of freedom  with frequency $\Omega$. The
vibronic coupling parameter is $\hbar g$. This system Hamiltonian
is assumed to be weakly coupled to right and left reservoirs of non-interacting
electrons (the baths), $H_{\mathrm{leads}}=\sum_{J\in R,L}\sum_{k\in J}\hbar\omega_{k}c_{k}^{\dagger}c_{k}$,
via the interaction term, $H_{\mathrm{int}}=\tilde{d}^{\dagger}\sum_{J\in R,L}\sum_{k\in J}\lambda_{k}c_{k}+h.c.$.
The relevant properties of the baths are encoded in the Fourier transforms
of the autocorrelation functions ($J=L,R$), $G_{J}(\omega)=\sum_{k\in J}\int_{-\infty}^{+\infty}e^{i\left(\omega-\omega_{k}\right)t}|\lambda_{k}|^{2}\langle c{}_{k}c_{k}^{\dagger}\rangle_{T_{J}}dt=e^{\hbar(\omega-\mu_{J})/k_{B}T_{J}}G_{J}(-\omega),$
where $\mu_{J}$ is the lead chemical potential, $k_{B}$ is the Boltzmann
constant and $T_{J}$ is the lead temperature. 

{}The analysis of this model is simplified by invoking the small
polaron transformation \cite{mahan2013many} which diagonalizes the
system Hamiltonian (see supplementary information), $H=\hbar\omega_{0}d^{\dagger}d+\hbar\Omega a^{\dagger}a,$
where $a,a^{\dagger}$ and $d,d^{\dagger}$ are transformed system
operators, and $\omega_{0}=\tilde{\omega}_{0}-\frac{g^{2}}{4\Omega}$.
{}We study the weak vibronic coupling limit of the above model, i.e.,
$\frac{g}{2\Omega}<<1$, which is realistic for many molecular systems
and was associated in earlier works with vibrational instability \cite{kochmorse,hartle2011vibrational,hartle2015effect}.
In this limit, each electron that flows through the junction exchanges
either one or zero vibration quanta with the bond oscillator. Our
purpose is to capture the onset of vibrational instability at low
excitation numbers, as the indicator for bond rupture in realistic
anharmonic systems. Therefore, we additionally restrict the following
analysis to low excitation numbers,

{} 
\begin{equation}
\frac{g}{2\Omega}\sqrt{N}<<1,\label{regime-1}
\end{equation}

\noindent {}where $N=\langle a^{\dagger}a\rangle$ is the average vibrational excitation. 

{}In a typical scenario where the coupling between the molecular
junction and the leads is weak, a Markovian master equation is adequate
for describing the evolution of $\rho_{n}^{0(1)}$  \cite{alicki4552periodically,davies1974markovian,gelbwaser2013work} which represents
the  population of the molecular electronic state and the vibrational
mode. The superscript $0(1)$ denotes a neutral (charged) electronic
level and $n$ the vibration quantum number. The master equation yields
the following equation of motion for the eigenstate populations, 
\begin{gather}
\dot{\rho}_{n}^{1}=G(-\omega{}_{0})\rho_{n}^{0}-G(\omega{}_{0})\rho_{n}^{1}+\nonumber \\
\frac{g^{2}}{4\Omega^{2}}\bigl\{ nG(-\omega_{+})\rho_{n-1}^{0}+(n+1)G(-\omega_{-})\rho_{n+1}^{0}\nonumber \\
-\bigl((1+n)G(\omega_{-})+nG(\omega_{+})\bigr)\rho_{n}^{1}\bigr\};\nonumber \\
\nonumber \\
\dot{\rho}_{n}^{0}=G(\omega{}_{0})\rho_{n}^{1}-G(-\omega{}_{0})\rho_{n}^{0}+\nonumber \\
\frac{g^{2}}{4\Omega^{2}}\bigl\{ nG(\omega_{-})\rho_{n-1}^{1}+(n+1)G(\omega_{+})\rho_{n+1}^{1}\nonumber \\
-\bigl((1+n)G(-\omega_{+})+nG(-\omega_{-})\bigr)\rho_{n}^{0}\bigr\},\label{eq:reddyn}
\end{gather}

\noindent {}where $\omega_{\pm}=\omega{}_{0}\pm\Omega$, and $G(\omega)=\sum_{J\in\{R,L\}}G_{J}(\omega)$. Generally, the coupling density to the $J$ lead depends on the temperature and the transition frequency, i.e., $G_{J}(\omega)=\Gamma_{J}(\omega)(1-f_{J}(\omega));G_{J}(-\omega)=\Gamma_{J}(\omega)f_{J}(\omega),$ where $\Gamma_{J}(\omega)$ is the rate of decay of the electronic occupation on the molecule due to molecule-lead coupling,  and $f_{J}(\omega)$ is the Fermi distribution \cite{peskin2010introduction}.

{}Notice that the  dynamics of the vibrational mode population
is affected by the leads through the electronic charging state. The weak electron-vibration coupling, renders this dynamics slow relatively
to the electronic evolution. This can be seen from Eqs. \eqref{eq:reddyn},
which point to two different time scales: The fast dynamics associated
with transfer between the electronic charging states, and the slow
dynamics of population transfer between vibrational states, which
depends on the small factor, $\frac{g^{2}}{\Omega^{2}}.$ 
 Accounting only for the fast dynamics, one obtains,

{} 
\begin{gather}
\dot{\rho}_{n}^{1}=-G(\omega{}_{0})\rho_{n}^{1}+G(-\omega{}_{0})\rho_{n}^{0};\nonumber \\
\dot{\rho}_{n}^{0}=-G(-\omega{}_{0})\rho_{n}^{0}+G(\omega{}_{0})\rho_{n}^{1},
\end{gather}

\noindent {}which implies that the electronic populations quickly
reach steady state, $\tilde{\rho}_{n}^{0(1)}=\tilde{\rho}_{n}\frac{G(\pm\omega_{0})}{G(\omega_{0})+G(-\omega_{0})}$,
where $\tilde{\rho}_{n}=\tilde{\rho}_{n}^{1}+\tilde{\rho}_{n}^{0}$
represents the population of the vibrational state $n$ after the
electronic states has reached steady state. Accounting also for the
terms proportional to $\frac{g^{2}}{\Omega^{2}}$, we now derive an
equation for $\tilde{\rho}_{n}$ on the slow time scale, which is
the equation of motion for the vibrational state populations, 

{} 
\begin{gather}
\dot{\tilde{\rho}}_{n}=r(n+1)\tilde{\rho}_{n+1}+sn\tilde{\rho}_{n-1}-(s(1+n)+rn)\tilde{\rho}_{n},\label{MEO-1}
\end{gather}

\noindent {}where $r(s)$ are the cooling (heating) rates,

{} 
\begin{gather}
r=\left(\frac{g}{2\Omega}\right)^{2}\frac{G(\omega_{+})G(-\omega{}_{0})+G(-\omega_{-})G(\omega{}_{0})}{G(\omega_{0})+G(-\omega{}_{0})};\nonumber \\
s=\left(\frac{g}{2\Omega}\right)^{2}\frac{G(\omega_{-})G(-\omega{}_{0})+G(-\omega_{+})G(\omega{}_{0})}{G(\omega{}_{0})+G(-\omega{}_{0})}.\label{eq:rands}
\end{gather}

\noindent {}These rates are composed of specific contributions. For
example, the product $G_{R}(\omega_{+})G_{L}(-\omega{}_{0})$ which
is included in the first term in $r$, corresponds to a cooling process
in which an electron with energy $\hbar\omega{}_{0}$ is being absorbed
from the left lead, followed by its emission to the right lead at
a different energy, $\hbar(\omega{}_{0}+\Omega)$. The net result
is a deexcitation of the vibrational mode by one quantum, i.e., $\hbar \Omega$.

{}The equation of motion for the average excitation energy, $\langle n(t)\rangle\equiv\sum_{n=1}^{\infty}n\rho_{n}(t)$,
can be readily obtained from Eq. \ref{MEO-1} , $\langle\dot{n}(t)\rangle=-r\langle n(t)\rangle+s(\langle n(t)\rangle+1)$,
which yields, 
\[
\langle n(t)\rangle=\frac{s}{r-s}+[\langle n(0)\rangle-\frac{s}{r-s}]e^{-(r-s)t}.
\]
In the scenario $r>s$, where the overall cooling rate exceeds the
overall heating rate, the vibrational mode reaches a stationary state
characterized by the asymptotic average  excitation, 
\begin{gather}
\langle n(\infty)\rangle=\frac{s}{r-s}.\label{eq:stat}
\end{gather}

{}Recalling that in realistic systems large excitation numbers would
be associated with highly anharmonic parts of the potential energy
surface, where bond rupture is likely to occur, we set a vibrational
excitation threshold level, $n_{\mathrm{tr}}$, beyond which the bond is considered
unstable. The condition for bond instability thus reads
$\frac{s}{r-s}>n_{\mathrm{tr}}$. Notice that this bounds from above the value
of $r-s.$ Hence, large $r-s$ values imply stable molecules, as suggested
by the fact that the overall cooling rates is larger than the overall
heating rate. Since the time evolution of $\langle n(t)\rangle$ is
monotonic, it is enough to consider the steady state in order to find
out if the vibrational mode population ever crossed the instability
threshold. When the overall heating rate exceeds the overall cooling
rate one has, $r<s$. Rather than approaching a steady state, the
vibrational excitation level diverges, implying that the junction
will be unstable for any $n_{\mathrm{tr}}$. This regime has been previously
related with work extraction in the context of heat machines  \cite{gelbwaser2013work,gelbwaser2015chapter}.

{} 
\begin{figure}
\begin{centering}
{}\includegraphics[width=0.48\textwidth]{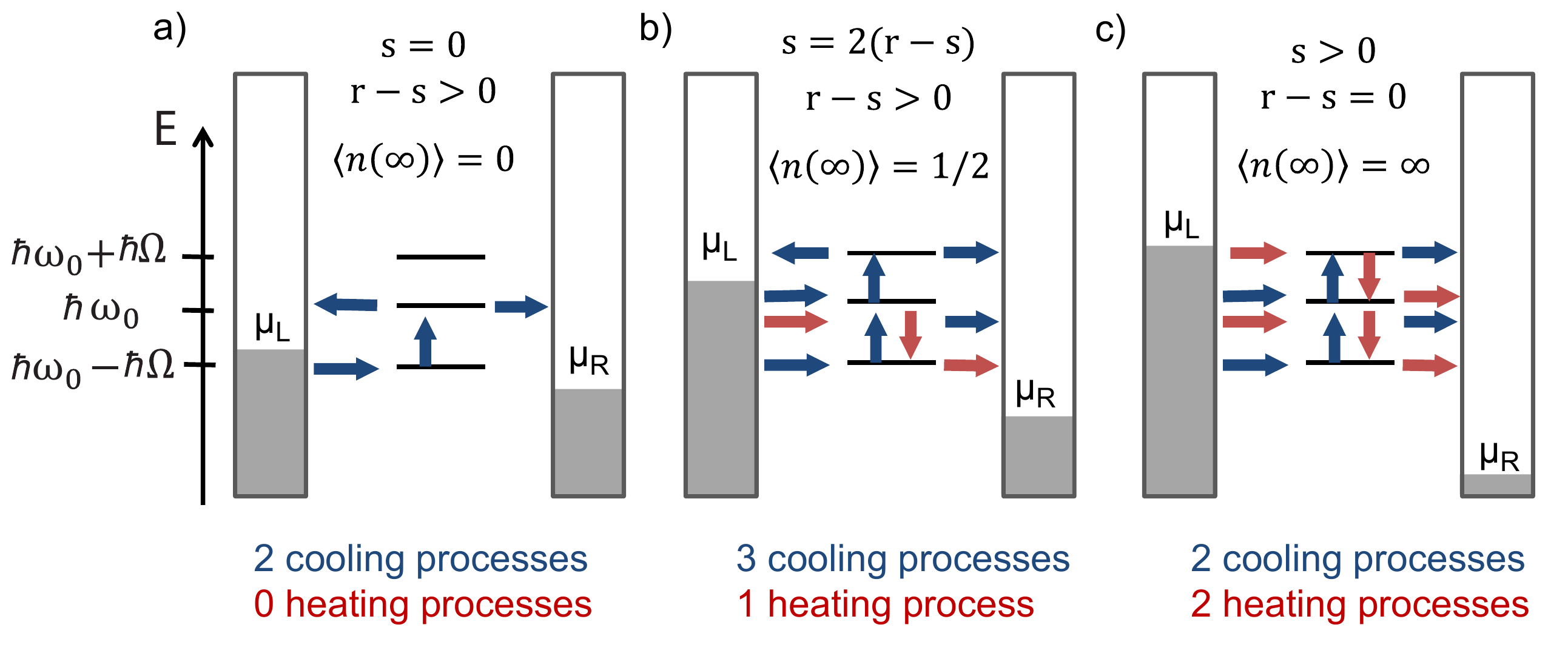} 
\par\end{centering}
{}\caption{Voltage-dependent heating (red) and cooling (blue) processes in the
wide-band limit for leads at zero temperature: a) low voltage; b)
intermediate voltage and c) high voltage.}
\label{pro} 
\end{figure}

{}Let us consider first the wide-band limit for the couplings to
the leads. The wide-band approximation implies that the energy-dependence
of the coupling densities $\{G_{J}(\omega)\}$ is only due to the
thermal electronic population in the lead, i.e., $G_{J}(\omega)\equiv\Gamma_{J}(1-f_{J}(\omega));G_{J}(-\omega)\equiv\Gamma_{J}f_{J}(\omega),$
where $\Gamma_{J}$ is a frequency-independent  decay rate \cite{peskin2010introduction}. 
In the zero temperature limit, $G_{J}(\omega)$ can take one of two values,
i.e., either zero or $\Gamma_{J}$ depending on the chemical potential.
Consequently, and for a symmetric junction with $\Gamma_{R}=\Gamma_{L}=\Gamma,$
the task of calculating the steady state excitation, $\frac{s}{r-s}$,
simplifies to counting the number of non-zero contributions to the
heating and cooling rates in Eqs. \eqref{eq:rands}. Fig. 1 depicts
schematically the three relevant scenarios, where the chemical potential
at the left lead is higher than that at the right lead. A larger bias
window, $\mu_{L}-\mu_{R}$, leads to an excess of heating over cooling
processes, which is reflected in a larger vibrational excitation number,
$\langle n(\infty)\rangle=0,\frac{1}{2},\infty$.

\begin{figure}[h]
\begin{centering}
{}\centering \includegraphics[width=0.5\textwidth]{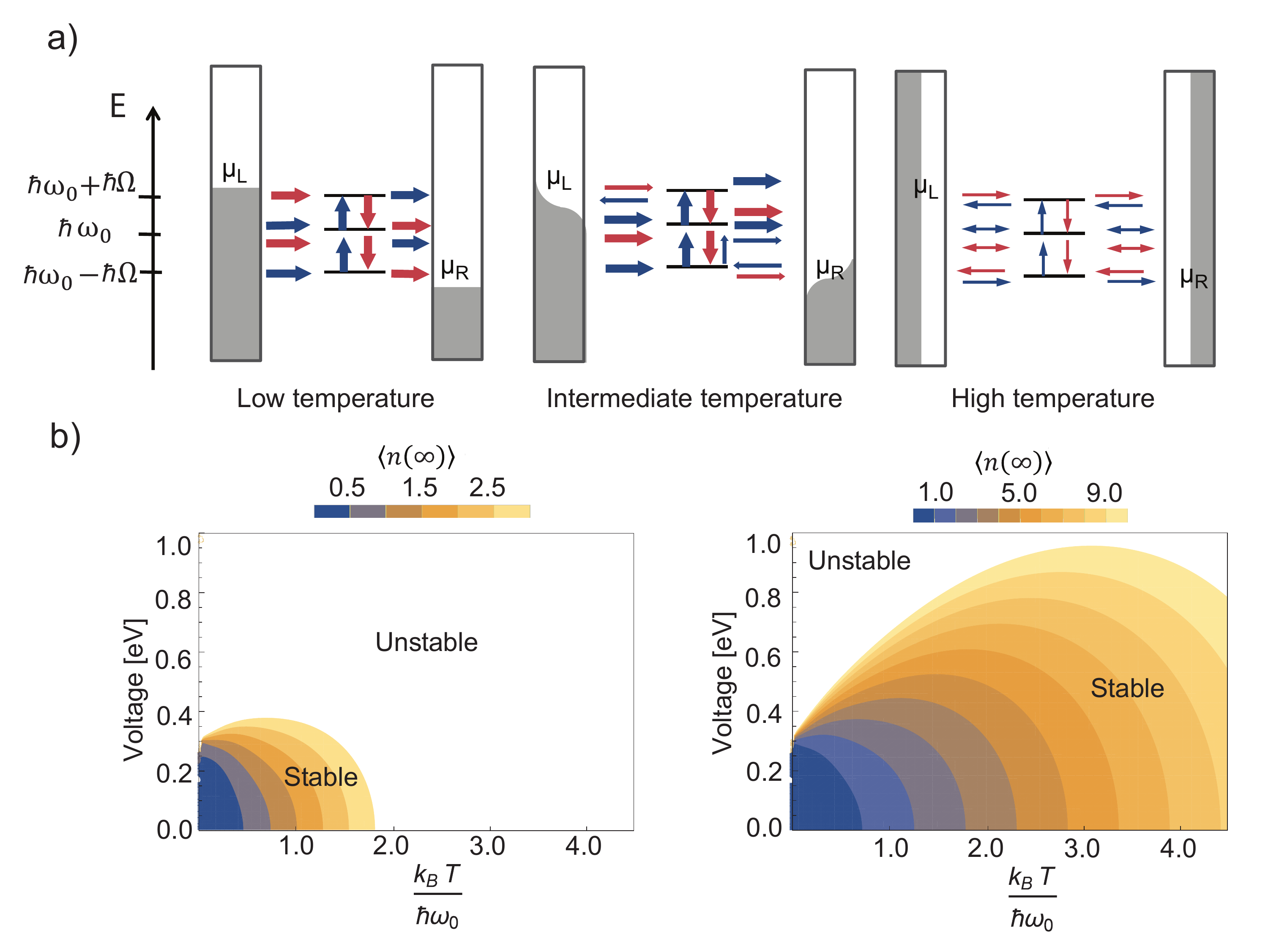} 
\par\end{centering}
\caption{a) Heating and cooling processes at high voltage for low, intermediate
and high temperatures (from left to right). b) Steady state vibrational
excitation, $\langle n(\infty)\rangle$, as function of the temperature
and the voltage in the wide-band limit. Colored areas represent 
regions of stability in contrast to white areas corresponding to bond instability.
The stability regions depends on $n_{\mathrm{tr}}$. The threshold is set to
$n_{\mathrm{tr}}=3$ on the left and to $n_{\mathrm{tr}}=10$ on the right. Notice the
different color scale between the two figures. The junction model
parameters are: $\hbar\omega_{0}=0.1$ eV, $\hbar\Omega=0.05$ eV,
$g=0.1\Omega$ and $\Gamma=0.01$ eV. }
\label{fig:temp}
\end{figure}

{}The trend in $\langle n(\infty)\rangle$, within the wide-band approximation,
seems to be in accord with recent experiments at finite (non-zero)
temperatures \cite{latha-si,capozzi2016mapping}, in which increasing the bias voltage was found to
lead to bond rupture. This trend is indeed observed also at finite
temperature within the present model, as demonstrated in Fig. 2.

{}The left and right plots in Fig. 2b correspond to the same junction
parameters (see figure caption), with $n_{\mathrm{tr}}=3$, and $n_{\mathrm{tr}}=10,$
respectively. Associating the bond instability with $\langle n(\infty)\rangle>n_{\mathrm{tr}}$,
the uncolored regions reflect the regions of bond instability, where
by definition, a smaller $n_{\mathrm{tr}}$ corresponds to a larger instability
region. Notice in Fig. 2 that, for certain voltages, $\langle n(\infty)\rangle$
decreases and then increases as a function of temperature. This non-monotonic dependence is due to the fact that the thermal broadening
of the Fermi distribution affects differently the heating and the cooling
rates. For unstable junctions (see Fig. 2a), a small increase of the
temperature permits electron-hole cooling processes and primarily
reduces vibrational heating processes. For example, the emission of
low energy electrons from the molecule to the right lead is partially
blocked in this case, thus reducing the overall vibrational heating
rate, and lowering $\langle n(\infty)\rangle$. A larger increase
of the temperature affects also the cooling rate by, among other things, reducing also
the emission of high energy electrons, which contributes to a relative
increase of $\langle n(\infty)\rangle$. At infinite temperature,
the Fermi distribution approaches the value $1/2$ for any frequency
and therefore all processes are allowed and have the same rate. Since
the numbers of allowed heating and cooling processes are the same,
the overall heating and cooling rates become equal, and so $\langle n(\infty)\rangle=\infty$,
destabilizing the junction for any voltage \cite{hartle2011vibrational}.

{} 
\begin{figure}
\includegraphics[width=0.48\textwidth]{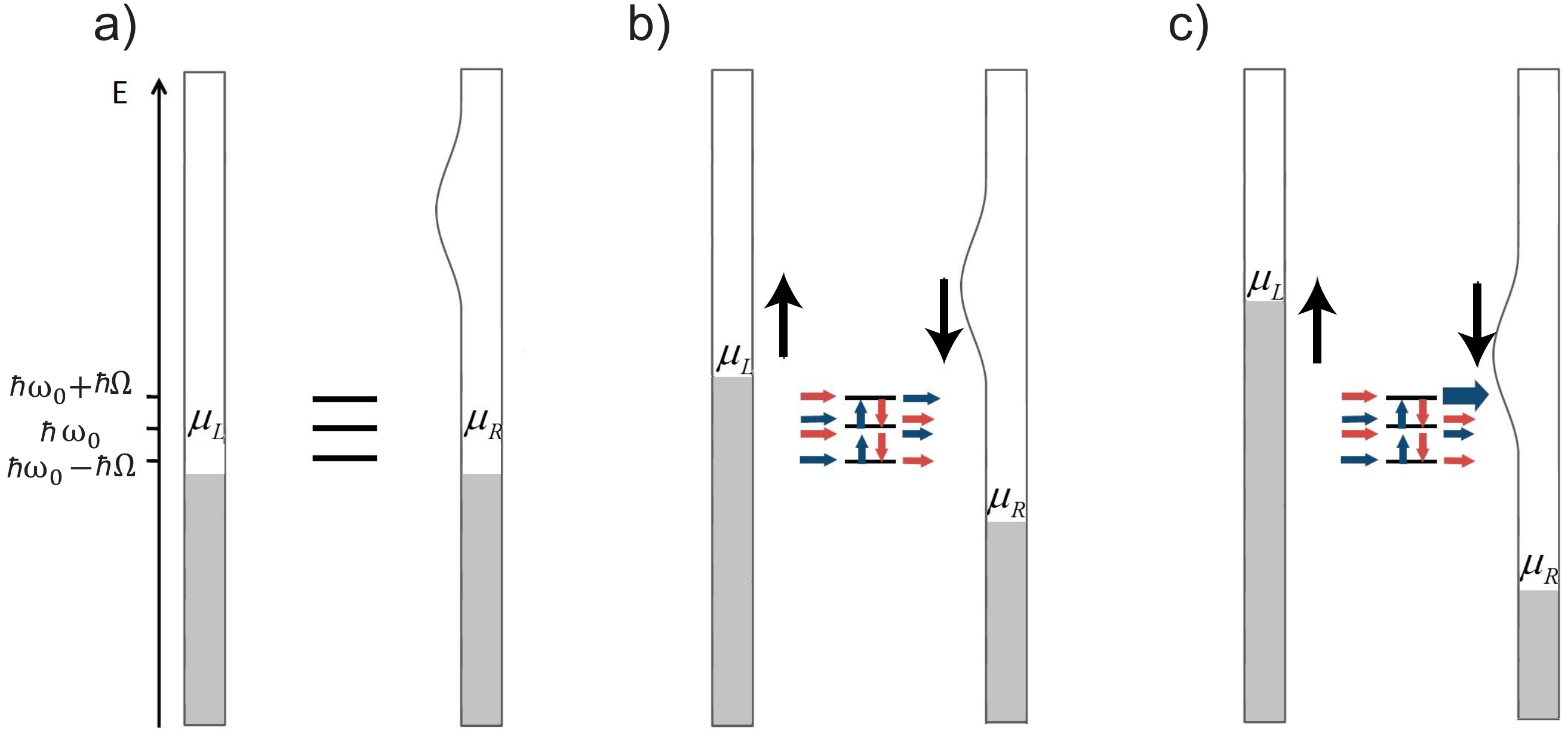} \caption{{\Large{}{}{}{}{}{}} Coupling spectrum beyond the wide-band approximation and its effect for zero (a), small (b)
and high (c) voltages. The voltage increase displaces the left and
right spectra in opposite directions. Therefore, the wide-band result
applies at low voltages, but breaks down at higher voltages, where
cooling processes becomes favorable with respect to heating.}
\label{fig:figure3} 
\end{figure}

{}A much richer voltage-dependence of the vibration instability is
expected in realistic systems where the assumption of the wide-band
approximation breaks down. While the wide-band approximation is often
adequate for describing the decay rates between molecules and metallic
leads, this approximation is an over simplification in other cases.
For example, graphene electrodes show rich energy-dependence of the
molecule-lead coupling, which depend on the particular graphene surface edge
coupled to the molecule \cite{ryndyk2012edge,ullmann2015single}. Even in the case
of metallic leads, covalently bonded adsorbates acting as linkers
between the metal and the conducting molecule may induce a pronounced
energy-dependence to the molecule-lead coupling. Accounting for the explicit
energy-dependence of the decay rates, $\Gamma_J\to\Gamma_{J}(\hbar\omega-\mu_{J})$,
the corresponding coupling densities obtain a non-trivial energy dependence
already at zero temperature. Since the latter determine rates of transport-induced
vibrational heating and cooling processes on the molecule, the lead
chemical potentials control in fact the balance between heating and
cooling and thus determine the bond stability in a non-trivial way.

{} 
\begin{figure}
{}\centering \includegraphics[width=0.38\textwidth]{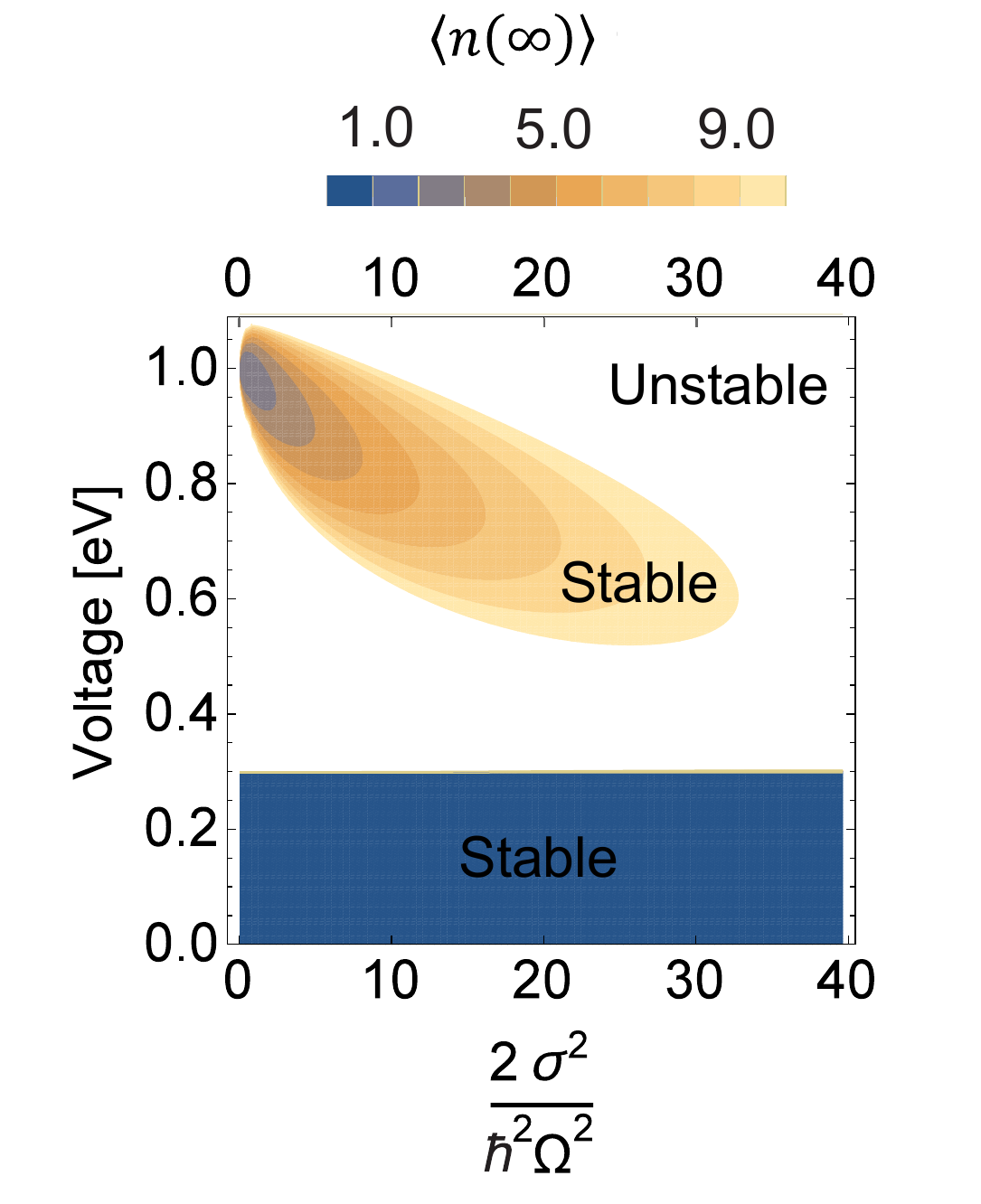}
\caption{{\footnotesize{}{}{}{}{}Steady state vibrational excitation, $\langle n(\infty)\rangle$,
beyond the wide-band approximation for a molecular junction at zero
temperature. Colored (uncolored) areas correspond to vibrational stability
(instability), where the threshold was set to $n_{\mathrm{tr}}=10$. The model
parameters are as in Fig. 2. The peak in the right lead decay rate above
the Fermi energy is centered around $\hbar\omega_{*}=0.65$ eV.}}
\label{fig:figure4} 
\end{figure}

{}Remarkably, in some realistic cases, an increase in the bias voltage
can actually stabilize the bond, in contrast to the intuitive result
based on the wide-band approximation. Without loss of generality,
let us consider a junction at zero temperature, where one of the leads
(the left one) has a flat electronic decay profile, $\Gamma_{L}(\hbar\omega-\mu_{L})=\Gamma$,
and the other (right) lead is also flat, except for an additional
Gaussian peak centered at $\mu_{R}+\hbar\omega_{*}$, i.e., $\Gamma_{R}(\hbar\omega-\mu_{R})=\Gamma\left(1+e^{-\frac{(\hbar\omega-\mu_{R}-\hbar\omega_{*})^{2}}{2\sigma^{2}}}\right)$,
as illustrated in Fig. 3a, where the leads Fermi energy is set to zero,
and the molecular charging energy is $\hbar \omega_{0}$. The presence of
an external bias voltage ($V$) on the junction is modeled here in
terms of shifts to the single particle energy levels in the non-interacting
leads, resulting in shifts of the molecule-lead coupling spectra (marked as vertical arrows in Figs. 3b, 3c). The left and right
chemical potentials become voltage-dependent ($\mu_{L}=eV/2$ and
$\mu_{R}=-eV/2$), and so do the electronic decay rates. As the voltage
increases, vibrational heating and cooling processes are activated.
If the vibration frequency is in the range, $\Omega\lesssim\omega_{*}-\omega_{0}-\sigma$
(see Fig. 3b), all the heating and cooling processes involving exchange
of a single vibration quanta become accessible at some voltage, while
the peak in $\Gamma_{R}(\hbar\omega-\mu_{R})$ is still outside the Fermi conductance window.
In this range, the model is in accord with the wide-band approximation
which predicts that a stable bond is destabilized by an increase of
the voltage (see Fig. 2a). This result is indeed confirmed also in
the lower part of Fig. 4 (V\textless{}0.4 eV). However, as the
voltage keeps increasing (see Fig. 3b), the peak in $\Gamma_{R}(\hbar\omega-\mu_{R})$ enters
the Fermi conductance window and the wide-band assumption no longer
holds. Consequently, the rate of vibrational cooling by electron emission
at energy $\hbar\omega=\hbar\omega_{0}+\hbar\Omega$ to the right
lead is favored over other processes. This breaking of the balance
between heating and cooling, leads to high-voltage induced stabilization
of the junction.

{}A more quantitative condition for the junction stability reads,
$\frac{s}{r-s}<n_{\mathrm{tr}}$. Using Eq. \eqref{eq:rands} for the overall
heating and cooling rates, this condition translates to the following
one,

\begin{equation}
\overline{\frac{d\Gamma_R}{d\omega}}>\frac{\overline{\Gamma}_R}{\Omega n_{tr}},\label{eq:condspec}
\end{equation}

\noindent {}where $\overline{\frac{d\Gamma_R}{d\omega}}\equiv\frac{1}{2\Omega}\int_{\omega_0-\Omega}^{\omega_0+\Omega}\frac{d\Gamma_R}{d\omega}d\omega$
is the average derivative and $\overline{\Gamma}_R=\frac{\Gamma_R(\hbar\omega_{0}-\mu_{r})+\Gamma_R(\hbar\omega_0-\hbar\Omega-\mu_{r})}{2}$
is the average electronic decay rate to the right lead. As shown on
Eq. \eqref{eq:condspec}, as long as the peak is steep enough, the
junction will be stabilized by increasing the bias voltage. 
This can be seen also from Fig. 4, where the color region shows
regimes where the junction is stable and the white area corresponds
to unstable junctions. Below a certain threshold, $\frac{2\sigma^{2}}{(\hbar\Omega)^{2}}\sim30$,
the bond is stabilized by increasing the voltage.

{}The above example demonstrates a general scenario of high voltage
induced mechanical stability, which is facilitated by a non-uniform energy dependence of the electron transfer rate between the molecule and the leads. Peaks (and deeps) in the transfer rate profiles
which characterize realistic lead structures and/or chemical compositions facilitate 
such a scenario. For example, employing graphene electrodes or specifically designed molecule-lead linker
groups, may be used to design mechanically stable single molecule devices operating
at high voltage in the resonance transport regime.

\begin{acknowledgements} This research was supported by the German
Israeli Foundation grant 1154-119.5/1. We acknowledge the support from the Center for
Excitonics, an Energy Frontier Research Center funded by the
U.S. Department of Energy under award DE-SC0001088 (Energy transfer).
UP acknowledges the great sabbatical
hospitality by the Harvard group. MT thanks Rainer {H\"artle} for helpful discussions.
\end{acknowledgements}

%\bibliographystyle{apsrev}
%\bibliography{bibhighv2}

\newpage
\onecolumngrid

\section*{Supplementary information}
\setcounter{equation}{0}
\renewcommand{\theequation}{S\arabic{equation}}

\setcounter{figure}{0}
\renewcommand{\thefigure}{S\arabic{figure}}

In this supplementary material we derive  the dynamic equations for the single molecular orbit, E, and the vibrational mode. We start by  defining new operators that diagonalize their Hamiltonian, which is
\begin{equation}
H=[\hbar\tilde{\omega}_{0}+\frac{\hbar g}{2}(\tilde{a}+\tilde{a}{}^{+})]\tilde{d}^{\dagger}\tilde{d}+\hbar\Omega\tilde{a}{}^{\dagger}\tilde{a}.
\label{eq:s1}
\end{equation}

The new operators  are  given by 

\begin{equation}
\tilde{a}\mapsto a=U^{\dagger}\tilde{a}U,\ \tilde{d}\mapsto d=U^{\dagger}\tilde{d}U; \quad U=e^{\frac{g}{2\Omega}(\tilde{a}^{\dagger}-\tilde{a})\tilde{d}^{\dagger}\tilde{d}}.
\label{eq:s2}
\end{equation}
With this new variables, the  Hamiltonian is diagonal,

\begin{equation}
H=\hbar\omega_{0}d^{\dagger}d+\hbar\Omega a^{\dagger}a,
\label{eq:s3}
\end{equation}
where $\omega_{0}=\tilde{\omega}_{0}-\frac{g^{2}}{4\Omega}$.  

We assume that E is weakly coupled to the leads therefore the reduced dynamics is governed by a Markovian master equation \cite{davies1974markovian,alicki4552periodically,gelbwaser2015chapter}. Furthermore, we analyze the regime where the vibrational mode is weakly coupled to E, $\frac{g}{\Omega}\ll 1$. Under these conditions, $\langle a^{\dagger}a\rangle\approx\langle\widetilde{a}^{+}\widetilde{a}\rangle$,
the transformed number operator $\langle a^{\dagger}a\rangle$
provides a reliable measure of the excitation level of the vibrational
mode in the original frame.

 In terms of the new operators the transformation is 

\begin{equation}
U=U^{\dagger}UU=e^{\frac{g}{2\Omega}U^{\dagger}(\tilde{a}^{\dagger}-\tilde{a})\tilde{d}^{\dagger}\tilde{d}U}=
e^{\frac{g}{2\Omega}(a^{\dagger}-a)d^{\dagger}d}
\label{eq:s4}
\end{equation}

and

\begin{equation}
\tilde{d}^{\dagger}=d^{\dagger}e^{-\frac{g}{2\Omega}(a^{\dagger}-a)}.
\label{eq:s5}
\end{equation}

We continue the derivation of the master equation by transforming the  operators of E in the interaction Hamiltonian, $\tilde{d}$ and $\tilde{d}^{\dagger}$, to the interaction picture:

\begin{equation}
\tilde{d}^{\dagger}(t)=e^{i H t} \tilde{d}^{\dagger}e^{-i H t} =e^{i\omega_{0}t}d^{\dagger}e^{\frac{g}{2\Omega}(a^{\dagger}e^{i\Omega t}-a e^{-i\Omega t})}\approx d^{\dagger}e^{i\omega{}_{0}t}+\frac{g}{2\Omega}\bigl(S_{1}^{\dagger}e^{i(\omega{}_{0}+\Omega)t}-S_{-1}^{\dagger}e^{i(\omega{}_{0}-\Omega)t}\bigr),
\label{eq:s6}
\end{equation}
where $S_{1}^{\dagger}=d^{\dagger}a^{\dagger}\ ,S_{-1}^{\dagger}=d^{\dagger}a$.
This approximation is strictly valid only for 
\begin{equation}
\frac{g}{2\Omega}\sqrt{\langle a^{\dagger}a\rangle}<<1.\label{sregime-1}
\end{equation}

Finally,  in the interactino picture the  dynamic equations for $\rho$, the density matrix of  E + vibrational mode, are

\begin{equation}
\dot{\rho}=\sum_{J\in R,L} \mathcal{L}_{0,J}+\mathcal{L}_{1,J}+\mathcal{L}_{-1,J},
\label{eq:s7}
\end{equation}
where

\begin{gather}
\mathcal{L}_{0,J} \rho=\frac{1}{2}  \left\{G_J(\omega_0)\left(
\left[d\rho,d^{\dagger}\right] +
\left[d,\rho d^{\dagger}\right]
\right)+
G_J(-\omega_0)\left(
\left[d^{\dagger}\rho,d\right] +
\left[d^{\dagger},\rho d\right]
\right)
\right\};  \notag\\
\mathcal{L}_{q,J} \rho=\frac{1}{2}  \left\{G_J(\omega_0+q\Omega)\left(
\left[S_q\rho,S_q^{\dagger}\right] +
\left[S_q,\rho S_q^{\dagger}\right]
\right)+
G_J(-(\omega_0+q\Omega))\left(
\left[S_q^{\dagger}\rho,S_q\right] +
\left[S_q^{\dagger},\rho S_q\right]
\right)
\right\}.
\label{eq:s8}
\end{gather}
\end{document}